\documentclass[12pt]{article}
\usepackage{amsmath}
\usepackage{amsfonts}
\usepackage{amssymb}
\usepackage{bm,mathtools}
\newcommand{\sgn}{\operatorname{sgn}}
\newcommand{\Tr}{\operatorname{Tr}}
\usepackage{graphicx}
\usepackage[nocompress]{cite}
\usepackage[hidelinks]{hyperref}
\hypersetup{pdfauthor={Lihong V. Wang},pdftitle={Derivations of Bloch (Majorana--Bloch) equation, von Neumann equation, and Schrodinger--Pauli equation}}
\title{Derivations of Bloch (Majorana--Bloch) equation, von Neumann equation, and Schrödinger--Pauli equation}
\author{Lihong V. Wang\\[0.45em]
\begin{minipage}{\dimexpr\textwidth-2\tabcolsep\relax}
\centering\fontsize{10}{12}\selectfont
Department of Electrical Engineering\\
Andrew and Peggy Cherng Department of Medical Engineering\\
California Institute of Technology\\
1200 E. California Blvd., MC 138-78, Pasadena, CA 91125, USA\\[0.25em]
\href{mailto:LVW@caltech.edu}{LVW@caltech.edu}\\
ORCID\quad\href{https://orcid.org/0000-0001-9783-4383}{0000-0001-9783-4383}
\end{minipage}
}
\date{Updated September 14, 2026}

\begin{document}

\maketitle

% Integration date September 13, 2026, America/Los_Angeles.
% Version date September 14, 2026, America/Los_Angeles.
% Revision E restores the manuscript content from revision C.
% The later review-related additions and their references are omitted.
% Supplied affiliations and contact details are retained from revision D.
% Equation boxes are removed; all numbered equations are preserved.
% The zero-induction limit, Lindblad comparison, and wording revisions remain.
% Appendix C source was created September 10, 2026 and revised September 11, 2026.
\begin{abstract} 
The transition from classical physics to quantum mechanics has been mysterious. Here, we derive the space-independent von Neumann equation for electron spin mathematically from the classical Bloch or Majorana--Bloch equation, which is also derived. Subsequently, the space-independent Schrödinger--Pauli equation is derived in both the quantum mechanical and recently developed co-quantum dynamic frameworks. An additional appendix incorporates the CQD signum branching function into the nonlinear von Neumann induction equation and develops its explicit double-commutator and state-dependent effective-Hamiltonian forms.
\end{abstract}

\setcounter{secnumdepth}{-1}
\section{Keywords}
\setcounter{secnumdepth}{3}
Bloch equation; Majorana--Bloch equation; von Neumann equation; Liouville–von Neumann equation; Schrödinger--Pauli equation; Schrödinger equation; Landau–Lifshitz–Gilbert equation; Electron spin

\section{Introduction} 
This manuscript is revised from Ref. \cite{wang2022derivation} by adding Appendix \ref{app:BE} --- Derivation of Bloch or Majorana--Bloch equation. The present update also adds Appendix~\ref{app:CQDnonlinear}, which combines the induction extension discussed in Section~\ref{sec:discussion} with the signum branching condition in Appendix~\ref{app:CQD} and derives the resulting nonlinear operator dynamics explicitly.

The Schrödinger equation, as a postulate, is a cornerstone in quantum mechanics. The transition from classical physics to quantum mechanics, however, remains a mystery. Various approaches to obtaining the time-dependent Schrödinger equation have been investigated \cite{Field2010,Scully2008,Briggs2007,Hall2002,Lamb2001,Lamb1994}. Recently, Schleich et al. generalized the Hamilton–Jacobi equation to reach the Schrödinger equation \cite{Schleich2013}. Most notably, Feynman used the path integral to attain the same equation \cite{Feynman1948}.

We investigate the transition from classical physics to the space-indepen\-dent Schrödinger--Pauli equation for electron spin. In classical electrodynamics, the motion of the magnetic dipole moment of an electron is governed by the Bloch (Majorana--Bloch) equation. Majorana stated that both the classical and the quantum-mechanical treatments on spin flip of atoms moving in a magnetic quadrupole field require integration of the same differential equations \cite{Majorana1932,Majorana2006}. It is known that the space-independent Schrödinger--Pauli equation or von Neumann equation (also known as the Liouville–von Neumann equation) for a unitary two-level system can be converted to the Bloch equation or its analog \cite{Grynberg2010,Feynman1963,Feynman1957}. However, the inverse conversion that would complete the two-way transitions has not been found in the literature.

Here, the classical Bloch equation for electron spin is mathematically converted to the space-independent von Neumann equation for a pure state of a two-level spin system. Subsequently, the space-independent Schrödinger--Pauli equation is derived in both frameworks of quantum mechanics and recently developed co-quantum dynamics (CQD, see Appendix \ref{app:CQD}). Therefore, the inverse conversion is shown, and the two-way transitions for a pure state of electron spin between the classical Bloch equation and the space-independent Schrödinger--Pauli equation are established.

\section{Derivation from Bloch equation to space-independent von Neumann equation}
\label{sec:bloch-vn}

We start with the classical Bloch equation derived in Appendix \ref{app:BE} for an electron in a magnetic field,
\begin{equation}
    \frac{d\vec{\mu}}{dt} = 
    \gamma \vec{\mu} \times \vec{B},
\end{equation}
where \(\vec{\mu}\) denotes the magnetic dipole moment of the electron, \(\gamma\) the gyromagnetic ratio, \(t\) time, and \(\vec{B}\) the magnetic flux density. Substitution of \(\vec{\mu} = \frac{\hbar}{2}\gamma\hat{\mu}\) yields
\begin{equation}
\frac{\hbar}{2} \frac{d\hat{\mu}}{dt} = \frac{\hbar}{2} \gamma \hat{\mu} \times \vec{B}.
\end{equation}
Here, \(\hbar\) denotes the reduced Planck constant, \(\frac{\hbar}{2}\) the spin angular momentum of the electron, and \(\hat{\mu}\) a unit vector. The unit vector is expressed as
\begin{equation}
\hat{\mu} = \begin{pmatrix}
\sin \theta \cos \phi \\
\sin \theta \sin \phi \\
\cos \theta
\end{pmatrix},
\end{equation}
where \(\theta\) and \(\phi\) denote the polar and azimuthal angles.

We now resort to the Pauli vector,
\begin{equation}
\vec{\sigma} = \sigma_x \hat{x} + \sigma_y \hat{y} + \sigma_z \hat{z},
\end{equation}
as a mathematical tool, which transforms real-space (\(x, y, z\)) vectors and operations using complex numbers. The Pauli matrices are given by
\begin{equation}
\sigma_x = \begin{pmatrix}0 & 1 \\ 1 & 0\end{pmatrix}, \quad \sigma_y = \begin{pmatrix}0 & -i \\ i & 0\end{pmatrix}, \quad \sigma_z = \begin{pmatrix}1 & 0 \\ 0 & -1\end{pmatrix}, \quad \sigma_0 = \begin{pmatrix}1 & 0 \\ 0 & 1\end{pmatrix}.
\end{equation}
Note that the Pauli matrices, related to quaternions, are applied beyond quantum mechanics. 

Multiplying both sides of the Bloch equation (Eq. 2) by \(\vec{\sigma}\) from the right yields
\begin{equation}
\frac{\hbar}{2} \left(\frac{d}{dt}\hat{\mu} \right)\cdot \vec{\sigma} = \frac{\hbar}{2} \gamma (\hat{\mu} \times \vec{B}) \cdot \vec{\sigma}.
\end{equation}
Merging the factors on the left-hand side and splitting the right-hand side produces
\begin{equation}
\frac{\hbar}{2} \frac{d}{dt} (\hat{\mu} \cdot \vec{\sigma}) = \frac{\hbar}{2} \gamma \left( \frac{1}{2} (\hat{\mu} \times \vec{B}) \cdot \vec{\sigma} - \frac{1}{2} (\vec{B} \times \hat{\mu}) \cdot \vec{\sigma} \right).
\end{equation}

Applying the following mathematical identity—known as the Pauli vector identity—for any two vectors \(\vec{a}\) and \(\vec{b}\),
\begin{equation}
(\vec{a} \cdot \vec{\sigma})(\vec{b} \cdot \vec{\sigma}) = (\vec{a} \cdot \vec{b}) \sigma_0 + i (\vec{a} \times \vec{b}) \cdot \vec{\sigma},
\end{equation}
we obtain
\begin{equation}
(\hat{\mu} \cdot \vec{\sigma})(\vec{B} \cdot \vec{\sigma}) = (\hat{\mu} \cdot \vec{B}) \sigma_0 + i (\hat{\mu} \times \vec{B}) \cdot \vec{\sigma}
\end{equation}
and
\begin{equation}
(\vec{B} \cdot \vec{\sigma})(\hat{\mu} \cdot \vec{\sigma}) = (\vec{B} \cdot \hat{\mu}) \sigma_0 + i (\vec{B} \times \hat{\mu}) \cdot \vec{\sigma}.
\end{equation}
Subtraction of the above two equations produces
\begin{equation}
(\hat{\mu} \cdot \vec{\sigma})(\vec{B} \cdot \vec{\sigma}) - (\vec{B} \cdot \vec{\sigma})(\hat{\mu} \cdot \vec{\sigma}) = 
i \left[ (\hat{\mu} \times \vec{B}) \cdot \vec{\sigma} -
(\vec{B} \times \hat{\mu}) \cdot \vec{\sigma} \right].
\end{equation}
This equation can be rewritten as \( \frac{1}{2} \left[(\hat{\mu} \cdot \vec{\sigma}), (\vec{B} \cdot \vec{\sigma})\right] = i (\hat{\mu} \times \vec{B}) \cdot \vec{\sigma} \).
Substituting into Eq. 7 yields
\begin{equation}
i \hbar \frac{d}{dt} 
\left( \frac{1}{2} \hat{\mu} \cdot \vec{\sigma} \right) = \frac{1}{2} \hbar \gamma 
\left( (\frac{1}{2} \hat{\mu} \cdot \vec{\sigma}) (\vec{B} \cdot \vec{\sigma}) - (\vec{B} \cdot \vec{\sigma}) 
(\frac{1}{2} \hat{\mu} \cdot \vec{\sigma}) \right).
\end{equation}

We define
\begin{equation}
\rho = \frac{1}{2} (\hat{\mu} \cdot \vec{\sigma} + \sigma_0).
\end{equation}
Adding \(\sigma_0\) above ensures that \(\rho\) has a unit trace.

Substituting Eq. 3–5 into Eq. 13 produces

\begin{equation}
\rho = \begin{pmatrix}
\cos^2 \frac{\theta}{2} & \cos \frac{\theta}{2} \sin \frac{\theta}{2} \exp(-i \phi) \\
\cos \frac{\theta}{2} \sin \frac{\theta}{2} \exp(i \phi) & \sin^2 \frac{\theta}{2}
\end{pmatrix},
\end{equation}
which reproduces the familiar quantum mechanical density matrix for a pure state of electron spin. One may express the density matrix using the state vector in outer-product form as follows:
\begin{align}
\rho = & \begin{pmatrix}
\cos \frac{\theta}{2} \\
\sin \frac{\theta}{2} \exp(i \phi)
\end{pmatrix}
\otimes
\begin{pmatrix}
\cos \frac{\theta}{2} \\
\sin \frac{\theta}{2} \exp(i \phi)
\end{pmatrix} \notag \\
= & \begin{pmatrix}
\cos \frac{\theta}{2} \\
\sin \frac{\theta}{2} \exp(i \phi)
\end{pmatrix}
\cdot
\begin{pmatrix}
\cos \frac{\theta}{2} & \sin \frac{\theta}{2} \exp(-i \phi)
\end{pmatrix}.
\end{align}

We next define
\begin{equation}
H = -\frac{1}{2} \hbar \gamma \vec{B} \cdot \vec{\sigma},
\end{equation}
which reproduces the familiar quantum mechanical Hamiltonian for electron spin. Note that \(\frac{1}{2} \hbar \gamma\) is the magnitude of the magnetic dipole moment.

Substituting Eq. 13 and 16 into Eq. 12 yields
\begin{equation}
i \hbar \frac{d}{dt} (\rho - \frac{1}{2} \sigma_0) = H (\rho - \frac{1}{2} \sigma_0) - (\rho - \frac{1}{2} \sigma_0) H.
\end{equation}
Eliminating the identity matrix \(\sigma_0\) (Eq. 5) produces the von Neumann equation,
\begin{equation}
i \hbar \frac{d}{dt} \rho = H \rho - \rho H = [H, \rho].
\end{equation}
Therefore, the classical Bloch equation is mathematically converted to the space-independent von Neumann equation for a pure state of electron spin.

\section{Quantum mechanical derivation from von Neumann equation to space-independent \\
Schrödinger--Pauli equation}
While the Schrödinger equation naturally evolves to the von Neumann equation \cite{Dirac1981}, the inverse process holds for a pure state \cite{Wieser2016}. The quantum mechanical density matrix for a pure state is given by
\begin{equation}
\rho = |\hat{\mu}\rangle\langle\hat{\mu}|.
\end{equation}
The ket and bra vectors \cite{Dirac1981} are given by
\begin{equation}
|\hat{\mu}\rangle = \begin{pmatrix}
\cos \frac{\theta}{2} \\
\sin \frac{\theta}{2} \exp(i \phi)
\end{pmatrix}
\end{equation}
and
\begin{equation}
\langle\hat{\mu}| = \begin{pmatrix}
\cos \frac{\theta}{2} &
\sin \frac{\theta}{2} \exp(-i \phi)
\end{pmatrix}.
\end{equation}

Substituting Eq. 19 into Eq. 18 yields
\begin{equation}
i \hbar \frac{d}{dt} \left(|\hat{\mu}\rangle\langle\hat{\mu}|\right) = \left[H, |\hat{\mu}\rangle\langle\hat{\mu}|\right].
\end{equation}
Expansion produces
\begin{equation}
\left(i \hbar \frac{d}{dt} |\hat{\mu}\rangle\right) \langle\hat{\mu}| + |\hat{\mu}\rangle \left(i \hbar \frac{d}{dt} \langle\hat{\mu}|\right) = (H|\hat{\mu}\rangle)\langle\hat{\mu}| - |\hat{\mu}\rangle (\langle\hat{\mu}|H).
\end{equation}
Combining terms results in
\begin{equation}
\left(i \hbar \frac{d}{dt} |\hat{\mu}\rangle - H|\hat{\mu}\rangle\right) \langle\hat{\mu}| = |\hat{\mu}\rangle \left(-i \hbar \frac{d}{dt} \langle\hat{\mu}| - \langle\hat{\mu}|H\right).
\end{equation}
Multiplying both sides by \(|\hat{\mu}\rangle\) from the right and using \(\langle\hat{\mu}|\hat{\mu}\rangle = 1\) yields
\begin{equation}
\left(i \hbar \frac{d}{dt} |\hat{\mu}\rangle - H|\hat{\mu}\rangle\right) = |\hat{\mu}\rangle \left(-i \hbar \frac{d}{dt} \langle\hat{\mu}| - \langle\hat{\mu}|H\right) |\hat{\mu}\rangle.
\end{equation}

If the bra and ket vectors \cite{Dirac1981} originate from independent realizations, both sides must vanish for the equation to hold. Therefore, we reach
\begin{equation}
i \hbar \frac{d}{dt} |\hat{\mu}\rangle = H|\hat{\mu}\rangle,
\end{equation}
which is the space-independent Schrödinger--Pauli equation.

Conversely, we now assume that the bra and ket vectors originate from identical realizations and examine the consequence. Differentiating \(\langle\hat{\mu}|\hat{\mu}\rangle = 1\) yields
\begin{equation}
\left(\frac{d}{dt} \langle\hat{\mu} |\right)|\hat{\mu}\rangle + \langle\hat{\mu}| \left(\frac{d}{dt} |\hat{\mu}\rangle \right) = 0.
\end{equation}
Substituting into Eq. 25 and using Eq. 19 produces
\begin{equation}
(1 - \rho) \left(i \hbar \frac{d}{dt} |\hat{\mu}\rangle - H|\hat{\mu}\rangle\right) = 0.
\end{equation}
If the matrix, \(1 - \rho\), is full rank, multiplying both sides from the left by the inverse matrix uniquely yields the space-independent Schrödinger--Pauli equation. However, from Eq. 14, we have the determinant, \(|1 - \rho| = 0\); thus, the matrix is singular. Consequently, the space-independent Schrödinger--Pauli equation can be reached as only a sufficient but not necessary condition. To reach the Schrödinger--Pauli equation as a sufficient and necessary condition, we assume that the bra and ket vectors \cite{Dirac1981} originate from independent realizations.

\section{CQD derivation from von Neumann \\equation to space-independent \\Schrödinger--Pauli equation}
In Section 3, the independent realizations are implicit in the bra and ket vectors (Eq. 20 and 21). On the basis of explicit independent realizations in CQD (see Appendix), we repeat the derivation. Henceforth, subscripted \(e\) and \(n\) denote the electron and nucleus, respectively, in the same atom.

The CQD pre-collapse state function is denoted by \(|\hat{\mu}_e \copyright \hat{\mu}_n\rangle\), where the co-quantum, \(\hat{\mu}_n\), is prefixed with \(\copyright\) for clarity. \(|\hat{\mu}_e \copyright \hat{\mu}_n\rangle\) represents the principal quantum, \(\hat{\mu}_e\), accompanied with \(\hat{\mu}_n\).

For a given \(\hat{\mu}_e\), the CQD prediction expressions for two independent realizations are written in dual spaces as follows:
\begin{equation}
|\hat{\mu}_e \copyright \hat{\mu}_{n1}\rangle = C_{1+}(\hat{\mu}_e, \hat{\mu}_{n1})|+z\rangle + C_{1-}(\hat{\mu}_e, \hat{\mu}_{n1})\exp (+i\phi_e)|-z\rangle
\end{equation}
and
\begin{equation}
\langle \hat{\mu}_e \copyright \hat{\mu}_{n2}| = C_{2+}(\hat{\mu}_e, \hat{\mu}_{n2})\langle +z| + C_{2-}(\hat{\mu}_e, \hat{\mu}_{n2}) \exp(-i\phi_e) \langle -z|.
\end{equation}
Numbers in subscripts denote independent realizations. Each binary coefficient represents either one or zero according to the CQD branching condition.

As shown in Section 2, the von Neumann equation was derived without ensemble averaging. Thus, we start with the following pre-averaging density operator:
\begin{equation}
\rho_0 = |\hat{\mu}_e \copyright \hat{\mu}_{n1}\rangle\langle\hat{\mu}_e \copyright \hat{\mu}_{n2}|.
\end{equation}
To illustrate the parallelism with Section 3, we keep the original wording as much as possible below.

The von Neumann equation becomes
\begin{equation}
i \hbar \frac{d}{dt} (|\hat{\mu}_e \copyright \hat{\mu}_{n1}\rangle\langle\hat{\mu}_e \copyright \hat{\mu}_{n2}|) = [H, |\hat{\mu}_e \copyright \hat{\mu}_{n1}\rangle\langle\hat{\mu}_e \copyright \hat{\mu}_{n2}|],
\end{equation}
where \(H\) is assumed to be shared by the two realizations. Expansion yields
\begin{align}
&\left( i \hbar  \frac{d}{dt} |\hat{\mu}_e \copyright \hat{\mu}_{n1}\rangle \right) \langle \hat{\mu}_e \copyright \hat{\mu}_{n2}| 
+ |\hat{\mu}_e \copyright \hat{\mu}_{n1}\rangle \left( i \hbar \frac{d}{dt} \langle \hat{\mu}_e \copyright \hat{\mu}_{n2}| \right) \notag \\
= &(H|\hat{\mu}_e \copyright \hat{\mu}_{n1}\rangle)\langle\hat{\mu}_e \copyright \hat{\mu}_{n2}| 
- |\hat{\mu}_e \copyright \hat{\mu}_{n1}\rangle(\langle\hat{\mu}_e \copyright \hat{\mu}_{n2}|H). 
\end{align}
Rearranging terms produces
\begin{align}
& \left( i \hbar \frac{d}{dt} - H \right) |\hat{\mu}_e \copyright \hat{\mu}_{n1}\rangle \langle \hat{\mu}_e \copyright \hat{\mu}_{n2}| \notag \\
= & |\hat{\mu}_e \copyright \hat{\mu}_{n1}\rangle \left( - i \hbar \frac{d}{dt} \langle \hat{\mu}_e \copyright \hat{\mu}_{n2}| - \langle \hat{\mu}_e \copyright \hat{\mu}_{n2}| H \right).
\end{align}
Multiplying both sides by \(|\hat{\mu}_e \copyright \hat{\mu}_{n2}\rangle\) from the right gives
\begin{align}
&\left[\left( i \hbar \frac{d}{dt} - H \right) |\hat{\mu}_e \copyright \hat{\mu}_{n1}\rangle \right] \langle \hat{\mu}_e \copyright \hat{\mu}_{n2}| \hat{\mu}_e \copyright \hat{\mu}_{n2}\rangle \notag \\
= &|\hat{\mu}_e \copyright \hat{\mu}_{n1}\rangle \left( - i \hbar \frac{d}{dt} \langle \hat{\mu}_e \copyright \hat{\mu}_{n2}| - \langle \hat{\mu}_e \copyright \hat{\mu}_{n2}| H \right) |\hat{\mu}_e \copyright \hat{\mu}_{n2}\rangle.
\end{align}

From Eq. 30, we have
\begin{align}
& \langle \hat{\mu}_e \copyright \hat{\mu}_{n2}| \hat{\mu}_e \copyright \hat{\mu}_{n2}\rangle \notag \\
= & C_{2+}^2 \langle +z|+z\rangle + C_{2-}^2 \langle -z|-z\rangle = C_{2+}^2 + C_{2-}^2 = 1.
\end{align}

Therefore, we reach
\begin{align}
& \left( i \hbar \frac{d}{dt} - H \right) |\hat{\mu}_e \copyright \hat{\mu}_{n1}\rangle \notag \\
= & |\hat{\mu}_e \copyright \hat{\mu}_{n1}\rangle \left( - i \hbar \frac{d}{dt} \langle \hat{\mu}_e \copyright \hat{\mu}_{n2}| - \langle \hat{\mu}_e \copyright \hat{\mu}_{n2}| H \right) |\hat{\mu}_e \copyright \hat{\mu}_{n2}\rangle,
\end{align}
which can be written alternatively as
\begin{align}
& \left( i \hbar \frac{d}{dt} - H \right) |\hat{\mu}_e \copyright \hat{\mu}_{n1}\rangle \notag \\
= & |\hat{\mu}_e \copyright \hat{\mu}_{n1}\rangle \left( \left( i \hbar \frac{d}{dt} - H \right) |\hat{\mu}_e \copyright \hat{\mu}_{n2}\rangle \right)^\dagger |\hat{\mu}_e \copyright \hat{\mu}_{n2}\rangle.
\end{align}
Because this equation holds for any two independent realizations, both sides must vanish, yielding
\begin{equation}
\left( i \hbar \frac{d}{dt} - H \right) |\hat{\mu}_e \copyright \hat{\mu}_{n1}\rangle = 0
\end{equation}
and
\begin{equation}
\left( i \hbar \frac{d}{dt} - H \right) |\hat{\mu}_e \copyright \hat{\mu}_{n2}\rangle = 0.
\end{equation}

Ensemble averaging either equation over all co-quantum realizations produces the space-independent Schrödinger--Pauli equation,
\begin{equation}
\left( i \hbar \frac{d}{dt} - H \right) |\hat{\mu}_e\rangle = 0.
\end{equation}

Conversely, we now assume that the bra and ket vectors originate from identical realizations and examine the consequence. Replacing each subscripted 2 with 1 in Eq. 37 yields
\begin{align}
& \left( i \hbar \frac{d}{dt} - H \right) |\hat{\mu}_e \copyright \hat{\mu}_{n1}\rangle \notag \\
= & |\hat{\mu}_e \copyright \hat{\mu}_{n1}\rangle \left( - i \hbar \frac{d}{dt} \langle \hat{\mu}_e \copyright \hat{\mu}_{n1}| - \langle \hat{\mu}_e \copyright \hat{\mu}_{n1}| H \right) |\hat{\mu}_e \copyright \hat{\mu}_{n1}\rangle.
\end{align}

Replacing each subscripted 2 with 1 in Eq. 36 gives
\begin{equation}
\langle \hat{\mu}_e \copyright \hat{\mu}_{n1}| \hat{\mu}_e \copyright \hat{\mu}_{n1}\rangle = 1.
\end{equation}
Differentiation yields
\begin{equation}
\left(\frac{d}{dt} \langle \hat{\mu}_e \copyright \hat{\mu}_{n1}|\right) |\hat{\mu}_e \copyright \hat{\mu}_{n1}\rangle + \langle \hat{\mu}_e \copyright \hat{\mu}_{n1}| 
\left(\frac{d}{dt} |\hat{\mu}_e \copyright \hat{\mu}_{n1}\rangle \right) = 0.
\end{equation}
Substitution into Eq. 42 gives
\begin{equation}
\left( i \hbar \frac{d}{dt} - H \right) |\hat{\mu}_e \copyright \hat{\mu}_{n1}\rangle = |\hat{\mu}_e \copyright \hat{\mu}_{n1}\rangle \langle \hat{\mu}_e \copyright \hat{\mu}_{n1}| \left( i \hbar \frac{d}{dt} - H \right) |\hat{\mu}_e \copyright \hat{\mu}_{n1}\rangle.
\end{equation}

Replacing each subscripted 2 with 1 in Eq. 31 and substituting it into the above equation produces
\begin{equation}
(1 - \rho_0) \left( i \hbar \frac{d}{dt} - H \right) |\hat{\mu}_e \copyright \hat{\mu}_{n1}\rangle = 0.
\end{equation}
If the matrix, \(1 - \rho_0\), is full rank, multiplying both sides from the left by the inverse matrix uniquely yields the space-independent Schrödinger--Pauli equation. However, from Eq. 14, which holds for \(\rho_0\) as well, we have the determinant, \(|1 - \rho_0| = 0\); thus, the matrix is singular. Therefore, the space-independent Schrödinger--Pauli equation can be reached as only a sufficient but not necessary condition. To reach the Schrödinger--Pauli equation as a sufficient and necessary condition, we assume that the bra and ket vectors \cite{Dirac1981} originate from independent realizations, which is explicit here.

\section{Discussion and summary}
\label{sec:discussion}
Quantum mechanics, celebrated for its countless triumphs, still poses open questions as discussed insightfully in recent monographs \cite{Bricmont2016,Norsen2017,Laloe2019,Auletta2019}. Various thought experiments have been proposed \cite{Einstein1935,Frauchiger2018,Schrodinger1935}. The transition from classical physics to quantum mechanics remains an open question. In the Copenhagen interpretation, an electron spin is considered to be simultaneously in both eigenstates, and its wavefunction collapses statistically upon measurement to either eigenstate \cite{Feynman1963}. The collapse of wavefunction is stated separately as a measurement postulate because it cannot be modeled by the original Schrödinger equation \cite{Norsen2017}.

We now extend the Bloch equation to the Landau–Lifshitz–Gilbert equation \cite{Gilbert2004},
\begin{equation}
\frac{d\hat{\mu}}{dt} = \gamma \hat{\mu} \times \vec{B} - k_i \hat{\mu} \times \frac{d\hat{\mu}}{dt}.
\label{eq:original-induction}
\end{equation}
Here, the dimensionless \(k_i\) is called the induction factor, which is used in CQD to explain the collapse of wavefunction (see Appendix). Although this equation was originally intended for condensed matter, the underlying physical mechanism for the added term is compatible with CQD. In fact, the author had developed CQD before realizing its connection with the Landau–Lifshitz–Gilbert equation. Setting \(k_i = 0\) recovers the Bloch equation.

Following the same procedure shown in Section 2, the Landau–Lifshitz– Gilbert equation (Eq. 47) can be converted to the following nonlinear variant of the von Neumann equation:
\begin{equation}
i \hbar \frac{d}{dt} \rho - \hbar k_i \left[ \frac{d\rho}{dt}, \rho \right] = [H, \rho].
\label{eq:original-nonlinear-vn}
\end{equation}
Appendix~\ref{app:CQDnonlinear} extends Eq.~\eqref{eq:original-nonlinear-vn} by inserting the explicit CQD signum branching factor. It derives the corresponding double-commutator form, expresses the branching sign in terms of the density operator under a prescribed constant nuclear polar angle, and constructs a Hermitian but state-dependent effective Hamiltonian. This extension retains a Hilbert-space representation without, in general, a single state-independent linear propagator for all initial states.

One may compare CQD with the existing quantum mechanical theories for collapse, e.g., the Ghirardi–Rimini–Weber model \cite{Ghirardi1986}, continuous spontaneous localization model \cite{Pearle1989,Ghirardi1990}, and the “Wavefunction Is the System Entity” (WISE) interpretation \cite{Long2021}.

Attempted conversion to a variant of the Schrödinger--Pauli equation has yielded only
\begin{equation}
i \hbar \frac{d}{dt} |\hat{\mu}\rangle - \hbar k_i (1 - \rho) \frac{d}{dt} |\hat{\mu}\rangle = H|\hat{\mu}\rangle,
\end{equation}
where \(\rho\) cannot be eliminated yet. This situation may be related to the fact that the collapse of wavefunction cannot be modeled by the original Schrödinger equation \cite{Norsen2017}.

In summary, the classical Bloch equation has been shown to lead to the space-independent von Neumann equation and the space-independent Schrödinger--Pauli equation, both for a pure state of electron spin. While it is known that the space-independent Schrödinger--Pauli equation or von Neumann equation for a unitary two-level system can be converted to the Bloch equation or its analog \cite{Grynberg2010,Feynman1963,Feynman1957}, the inverse conversion for electron spin is shown here. It is first shown that the Bloch equation and the space-independent von Neumann equation are equivalent for a pure state of electron spin. Further conversion from the space-independent von Neumann equation to the space-independent Schrödinger--Pauli equation as a both sufficient and necessary condition is proven under the assumption of independent realizations of the bra and ket vectors, which is implicit in quantum mechanics but explicit in CQD. Without such an assumption, the space-independent Schrödinger--Pauli equation is only a sufficient but not necessary condition to either the classical Bloch equation or the space-independent von Neumann equation. The presented transition from classical physics to quantum mechanics can potentially lead to new insight into some of the open questions.

\setcounter{secnumdepth}{-1}
\section{Acknowledgments}
\setcounter{secnumdepth}{3}
The author thanks Dr. Zhe He for discussing the manuscript and verifying the derivations, Prof. JT Shen for discussing the work, and Dr. Kelvin Titimbo and Suleyman Kahraman for discussing Appendix \ref{app:BE}. 

\setcounter{secnumdepth}{-1}
\section{Declarations}
\setcounter{secnumdepth}{3}
\textbf{Funding:} No funding was received for conducting this study.\\
\textbf{Conflicts of interest/Competing interests:} The authors have no relevant financial or non-financial interests to disclose.\\
\textbf{Availability of data and material:} All data used in this study are available from the author upon reasonable request.\\
\textbf{Code availability:} All custom codes used in this study are available from the author upon reasonable request.\\
 
\bibliographystyle{unsrt} 
%\bibliography{BE_submit_bib}

\appendix
\newpage
\section{Co-quantum dynamics} \label{app:CQD}
An essential excerpt on co-quantum dynamics (CQD)\cite{wang2023multiJPb, wang2023multiArXiv, titimbo2023numerical, he2023numerical} is provided below for completeness.

\begin{abstract}
In the classic multi-stage Stern–Gerlach experiment conducted by Frisch and Segrè, the Majorana (Landau–Zener) or Rabi formulae diverge afar from the experimental observation while the physical mechanism for electron-spin collapse remains unidentified. Here, introducing the physical co-quantum concept provides a plausible physical mechanism and predicts the experimental observation in absolute units without fitting (i.e., no parameters adjusted) highly accurately. Further, the co-quantum concept is corroborated by statistically reproducing exactly the wave function, density operator, and uncertainty relation for electron spin.
\end{abstract}

In typical Stern–Gerlach experiments, the dominant motion of \(\hat{\mu}_e\) is precession about the main field, and the secondary motion is collapse due to induction—with the following trend:
\begin{equation}
\tan \frac{\theta_e(t)}{2} = \tan \frac{\theta_e(0)}{2} \exp\left[-\text{sgn}(\theta_n - \theta_e) k_i |\Delta\phi_e(t)|\right].
\label{eq:original-cqd-collapse}
\end{equation}
Here, \(\Delta\phi_e\) denotes the traversed azimuthal angle (i.e., the phase). As time evolves, \(\theta_e\) approaches either 0 or \(\pi\) according to the following branching condition:
\begin{equation}
\text{sgn}(\theta_n - \theta_e) = 
\begin{cases}
1 & \text{if } \theta_n > \theta_e,\\
0 & \text{if } \theta_n = \theta_e,\\
-1 & \text{else}.
\end{cases}
\label{eq:original-cqd-branching}
\end{equation}
Therefore, \(\hat{\mu}_e\) collapses to either \(+z\) or \(-z\).

The CQD pre-collapse state function is denoted by \(|\hat{\mu}_e \copyright \hat{\mu}_n\rangle\), where the co-quantum, \(\hat{\mu}_n\), is prefixed with \(\copyright\) for clarity. \(|\hat{\mu}_e \copyright \hat{\mu}_n\rangle\) represents \(\hat{\mu}_e\) accompanied with \(\hat{\mu}_n\).

The CQD prediction expression for Stern–Gerlach experiments is written as
\begin{equation}
|\hat{\mu}_e \copyright \hat{\mu}_n\rangle = C_+(\hat{\mu}_e, \hat{\mu}_n)|+z\rangle + C_-(\hat{\mu}_e, \hat{\mu}_n) \exp(i\phi_e) | -z\rangle.
\end{equation}
The equal sign functions as a right arrow (\(\rightarrow\)) because the right-hand side predicts the measurement outcome. A given \(\hat{\mu}_e\) collapses to either \(+\hat{z}\) or \(-\hat{z}\) according to the branching condition (Eq. 51). The two real and positive \(C\) coefficients take on mutually exclusive binary values while \(\exp(i\phi_e)\) captures the phase information. If \(\theta_n > \theta_e\), then \(C_+ = 1\) and \(C_- = 0\); if \(\theta_n < \theta_e\), \(C_+ = 0\) and \(C_- = 1\). In either case, \(C_+ \cdot C_- = 0\) and \(C_+ + C_- = 1\).

\newpage
\section{Derivation of Bloch or Majorana--Bloch \\equation} \label{app:BE}
Majorana formulated the equation commonly known as the ``Bloch equation'' \cite{Majorana1932, Majorana2006} fourteen years before Bloch published his version in 1946 \cite{bloch1946nuclear}. Whereas Majorana applied the equation to an atom, Bloch asserted that it is valid only for macroscopic magnetization. The author proposes that the Bloch equation be renamed the Majorana--Bloch equation. Since the derivation of the Bloch equation is absent in both Majorana’s and Bloch’s works, as well as in the literature, to the best of our knowledge, we present our derivation for completeness using two models below. 

\subsection{Current-loop model}
We first use a current loop to model the angular momentum $\vec{S}$ and the magnetic moment $\vec{\mu}$. The Newton's second law in a rotational system states
\begin{equation}
\frac{d}{dt}\vec{S} = \vec{\tau}, \label{eq:dSdt=tau_loop}
\end{equation}
where $\vec{\tau}$ denotes the torque.

In the presence of an external magnetic field \(\vec{B}\), the differential torque due to the Lorentz force can be written as
\begin{align}
d\vec{\tau} &=  \vec{r} \times \left( I d\vec{r} \times \vec{B} \right)  \notag \\
     &= I  \left( \vec{r} \cdot \vec{B} \right) d\vec{r} - I \left( \vec{r} \cdot d\vec{r} \right) \vec{B}  \notag \\
     &= I  \left( \vec{r} \cdot \vec{B} \right) d\vec{r}, \label{eq:tau_loop}
\end{align}
where \(\vec{r}\)  denotes the position vector relative to the center of the loop, $I$ the current, and $d\vec{r}$  a differential segment of the loop. 

Without losing generality, we center-align the loop with the $xy$ plane and further align the $x$-axis such that $B_y = 0$. Using the azimuthal angle $\phi$, we express
\begin{equation}
\vec{r} = R \cos(\phi) \hat{x} + R \sin(\phi) \hat{y},
\end{equation}
where $R$ denotes the radius of the current loop. Consequently, we have
\begin{equation}
\vec{r} \cdot \vec{B} = R \cos(\phi) B_x
\end{equation}
and
\begin{equation}
d\vec{r} = -R \sin(\phi) d\phi \hat{x} + R \cos(\phi) d\phi \hat{y}.
\end{equation}

Substituting the preceding two equations into Eq. \ref{eq:tau_loop}, we  complete the integral $\vec{\tau} = \int d\vec{\tau}$ over the loop, yielding
\begin{equation}
\vec{\tau} =  \pi I R^2 \hat{z} \times \vec{B}.
\end{equation}

Note that the magnetic moment of a current loop is defined as the product of the current and the area enclosed by the loop, with the direction given by the right-hand rule (perpendicular to the plane of the loop):
\begin{equation}
\mu = I \cdot {\pi R^2}.
\end{equation}
Therefore, we obtain
\begin{equation}
\vec{\tau} = \vec{\mu} \times \vec{B}.
\end{equation}
Substitution of this equation into Eq. \ref{eq:dSdt=tau_loop} yields
\begin{equation}
\frac{d}{dt} \vec{S} = \vec{\mu} \times \vec{B}.
\end{equation}

Using the gyromagnetic ratio,
\begin{equation}
\gamma = \frac{\mu}{S},
\end{equation}
we obtain the Bloch or Majorana--Bloch equation:
\begin{equation}
\frac{d}{dt} \vec{\mu} = \gamma \vec{\mu} \times \vec{B}.
\end{equation}

Given the charge $q$ and mass $m$ rotating at the angular velocity $\omega$, we compute the gyromagnetic ratio classically as
\begin{equation}
    \gamma = \frac{\mu}{S} 
    = \frac{ I \cdot \pi R^2}{m R^2 \cdot \omega } 
    = \frac{\frac{1}{2} q \omega R^2}{m \omega R^2}  = \frac{q}{2m}.
\end{equation}

\subsection{Point-particle model}
We next consider a point-particle model. For comparison, we keep some of the above descriptions verbatim. The point mass \(m\) circles with velocity \(\vec{v}\),  subject to an intrinsic force $\vec{F_i}$ that governs the circular motion, in addition to an extrinsic force \(\vec{F_e}\). Newton’s second law states
\begin{equation}
m \frac{d}{dt} \vec{v} = \vec{F_i} + \vec{F_e}.
\end{equation}
Cross-multiplying the equation with the position vector of the point particle \(\vec{r}\) from the left-hand side yields
\begin{equation}
\vec{r} \times m \frac{d}{dt} \vec{v} = \vec{r} \times \vec{F_e},
\end{equation}
where $\vec{r} \times \vec{F_i} \approx 0$ because  $\vec{F_i} \parallel \vec{r} $ holds over one cycle $T$ if the circular motion is much faster than precession. 

When $\vec{F_e}$ is provided by an external magnetic field \(\vec{B}\) on the charge \(q\) of the point mass, substituting the Lorentz force formula yields
\begin{equation}
\vec{r} \times m \frac{d}{dt} \vec{v} = \vec{r} \times \left( q \vec{v} \times \vec{B} \right).
\end{equation}
Adding the above equation with
\begin{equation}
\left(\frac{d}{dt}  \vec{r} \right) \times m \vec{v}  = \vec{v} \times m \vec{v} = 0,
\end{equation}
we update the left-hand side and reach
\begin{equation}
\frac{d}{dt} \left( \vec{r} \times m \vec{v} \right) 
= \vec{r} \times \left( q \vec{v} \times \vec{B} \right).
\end{equation}
The left-hand side is related to the instantaneous angular momentum
\begin{equation}
    \vec{S_i} =  \vec{r} \times m \vec{v},
\end{equation}
and the right-hand side is the instantaneous torque
\begin{equation}
    \vec{\tau_i} =  \vec{r} \times \left( q \vec{v} \times \vec{B} \right).
\end{equation}
Consequently, we obtain the rewritten form:
\begin{equation}
\frac{d}{dt} \vec{S_i} 
= \vec{\tau_i}.
\end{equation}

We now average both sides of the equation over one cycle $T$:
\begin{equation}
\left\langle \frac{d}{dt} \vec{S_i} \right\rangle 
= 
\left\langle \vec{\tau_i}\right\rangle. \label{eq:DSi=taui_avg}
\end{equation} 

The left-hand side can be expressed as
\begin{align}
\left\langle \frac{d}{dt} \vec{S_i}(t) \right\rangle  
&= \frac{1}{T} \int_{t - T/2}^{t + T/2} \frac{d \vec{S_i}(\tau)}{\partial \tau} \, d\tau \notag \\ 
&=  \frac{1}{T} \left( \vec{S_i}(t + T/2) - \vec{S_i}(t - T/2) \right).
\end{align} 
A commuted form can be expressed below using the Leibniz integral rule:
\begin{align}
\frac{d}{dt} \langle \vec{S_i}(t) \rangle 
&= \frac{d}{dt} \left( \frac{1}{T} \int_{t - T/2}^{t + T/2} \vec{S_i}(\tau) \, d\tau \right) \notag \\
& = \frac{1}{T} \left( \vec{S_i}(t + T/2) - \vec{S_i}(t - T/2) \right).
\end{align}
Comparing the above two equations leads to
\begin{equation}
\left\langle \frac{d}{dt} \vec{S_i}(t) \right\rangle =
\frac{d}{dt} \langle \vec{S_i}(t) \rangle.
\end{equation}
We use \(\vec{S}\) to denote the average angular momentum $\langle \vec{S_i}(t) \rangle$. One can also derive the above commutation as follows:
\begin{align}
\left\langle \frac{d\vec{S}_i(t)}{dt} \right\rangle &= \left\langle \lim_{\Delta t \to 0} \frac{\vec{S}_i(t+\Delta t) - \vec{S}_i(t)}{\Delta t} \right\rangle \notag \\
&= \lim_{\Delta t \to 0} \frac{\langle \vec{S}_i(t+\Delta t) \rangle - \langle \vec{S}_i(t) \rangle}{\Delta t} \notag \\
&= \lim_{\Delta t \to 0} \frac{\vec{S}(t+\Delta t) - \vec{S}(t)}{\Delta t} \notag \\
&= \frac{d\vec{S}(t)}{dt}.
\end{align}

Substituting into Eq. \ref{eq:DSi=taui_avg} yields
\begin{equation}
\frac{d}{dt} \vec{S}(t) 
= 
\left\langle \vec{\tau_i}\right\rangle. \label{eq:DS=taui_avg}
\end{equation} 

We now evaluate the right-hand side of the above equation for the average torque, denoted by $\vec{\tau}$:
\begin{align}
\vec{\tau} &= \left\langle \vec{r} \times \left( q \vec{v} \times \vec{B} \right) \right\rangle \notag \\
     &= q \left\langle \left( \vec{r} \cdot \vec{B} \right) \vec{v} - \left( \vec{r} \cdot \vec{v} \right) \vec{B} \right\rangle \notag \\
     &= q \left\langle \left( \vec{r} \cdot \vec{B} \right) \vec{v} \right\rangle. \label{eq:tau}
\end{align}

Without losing generality, we center-align the circular trajectory with the $xy$ plane and further align the $x$-axis such that $B_y = 0$. The position vector \(\vec{r}\) versus time $t$ can be expressed as
\begin{equation}
\vec{r} = R \cos(\omega t) \hat{x} + R \sin(\omega t) \hat{y},
\end{equation}
where $R$ denotes the radius of the circular motion and $\omega$ the angular velocity. As the average is over one cycle, we choose $t$ such that the initial phase is $0$. Consequently, we have
\begin{equation}
\vec{r} \cdot \vec{B} = R \cos(\omega t) B_x
\end{equation}
and
\begin{equation}
\vec{v} = \frac{d\vec{r}}{dt} = -R \omega \sin(\omega t) \hat{x} + R \omega \cos(\omega t) \hat{y}.
\end{equation}

Substituting the preceding two equations into Eq. \ref{eq:tau} and completing the averaging over one cycle using
\begin{equation}
\langle \cdot \rangle =  \frac{1}{T} \int_{t - T/2}^{t + T/2} \cdot \; d\tau
\end{equation}
yield
\begin{equation}
\vec{\tau} =  \frac{1}{2} q v R \hat{z} \times \vec{B}.
\end{equation}

Note that the magnetic moment of the equivalent current loop is defined as the product of the current and the area enclosed by the loop:
\begin{equation}
\mu = \frac{q}{\frac{2\pi R}{v}} {\pi R^2} = \frac{1}{2} q v R.
\end{equation}
Therefore, we reach
\begin{equation}
\vec{\tau} = \vec{\mu} \times \vec{B}.
\end{equation}
Substitution of this equation into Eq. \ref{eq:DS=taui_avg} yields
\begin{equation}
\frac{d}{dt} \vec{S} = \vec{\mu} \times \vec{B}.
\end{equation}

Using the gyromagnetic ratio,
\begin{equation}
\gamma = \frac{\mu}{S},
\end{equation}
we obtain the Bloch or Majorana--Bloch equation:
\begin{equation}
\frac{d}{dt} \vec{\mu} = \gamma \vec{\mu} \times \vec{B}.
\end{equation}

Given the charge $q$ and mass $m$, we compute the gyromagnetic ratio classically as
\begin{equation}
    \gamma = \frac{\mu}{S} 
    = \frac{ \frac{1}{2} q v R}{m v R} 
    = \frac{q}{2m}.
\end{equation}

Here, we use circular motion to emulate spin. In contrast, in relativistic quantum mechanics, the momentum density in the Dirac wave field includes a term related to the circulation of energy in the rest frame of the electron \cite{ohanian1986spin}.

One can extend the above equation from a point particle to an object of other shapes, such as a loop or spheroid, rotating at $\omega$. The time evolution of a differential element is given by
\begin{equation}
\frac{1}{\gamma} \frac{d(d\vec{\mu})}{dt} = d\vec{\mu} \times \vec{B}.
\end{equation}
If $\gamma$ is uniform, integrating both sides over the entire object yields
\begin{equation}
\frac{1}{\gamma} \frac{d}{dt} \left( 
\int {d\vec{\mu}} \right) = \left( 
\int {d\vec{\mu}} \right) \times \vec{B}.
\end{equation}
Equating each integral to the total magnetic moment $\vec{\mu}$  yields the Bloch or Majorana--Bloch equation:
\begin{equation}
\frac{1}{\gamma} \frac{d\vec{\mu}}{dt} = \vec{\mu} \times \vec{B}.
\end{equation}
If $\gamma$ is nonuniform, one may define an effective value using a $g$ factor.

\clearpage
\section{Nonlinear von Neumann form of the CQD induction equation\\with the signum branching function}
\label{app:CQDnonlinear}
\setcounter{equation}{0}
\renewcommand{\theequation}{\thesection.\arabic{equation}}
\renewcommand{\theHequation}{\thesection.\arabic{equation}}

% Incorporated from CQD_nonlinear_von_Neumann_2026_09_11c.tex.
% Original creation date September 10, 2026.
% Source version date September 11, 2026.
% Integration date September 13, 2026, America/Los_Angeles.
% This appendix retains the source's eight-part derivation and equations.

\subsection{Starting CQD equation}

Consider the CQD extension of the Bloch (Majorana--Bloch) equation for the electron magnetic-moment unit vector \(\hat{\bm\mu}_e\). Incorporating explicitly the CQD branching sign gives
\begin{equation}
\frac{d\hat{\bm\mu}_e}{dt}
=
\gamma_e\hat{\bm\mu}_e\times\bm B
-
s k_i\hat{\bm\mu}_e\times
\frac{d\hat{\bm\mu}_e}{dt},
\label{eq:cqdvn-cqd-vector}
\end{equation}
where \(k_i\) is the dimensionless induction factor and
\begin{equation}
s\equiv \sgn(\theta_n-\theta_e)
=
\begin{cases}
+1, & \theta_n>\theta_e,\\
0, & \theta_n=\theta_e,\\
-1, & \theta_n<\theta_e.
\end{cases}
\label{eq:cqdvn-signum}
\end{equation}
The signum function selects the CQD collapse branch. In Appendix~\ref{app:CQD}, Eq.~\eqref{eq:original-cqd-collapse} gives the associated angular evolution
\begin{equation}
\tan\frac{\theta_e(t)}{2}
=
\tan\frac{\theta_e(0)}{2}
\exp\!\left[-s k_i\left|\Delta\phi_e(t)\right|\right].
\label{eq:cqdvn-collapse}
\end{equation}
Thus \(s=+1\) drives \(\theta_e\to0\), whereas \(s=-1\) drives \(\theta_e\to\pi\).

\subsection{Density operator and Hamiltonian}

Define, as in the Bloch-to-von-Neumann conversion in Section~\ref{sec:bloch-vn},
\begin{equation}
\rho=\frac{1}{2}\left(\sigma_0+\hat{\bm\mu}_e\cdot\bm\sigma\right),
\label{eq:cqdvn-rho}
\end{equation}
and
\begin{equation}
H=-\frac{\hbar\gamma_e}{2}\bm B\cdot\bm\sigma.
\label{eq:cqdvn-H}
\end{equation}
The Pauli-vector identity
\begin{equation}
(\bm a\cdot\bm\sigma)(\bm b\cdot\bm\sigma)
=(\bm a\cdot\bm b)\sigma_0
+i(\bm a\times\bm b)\cdot\bm\sigma
\end{equation}
implies
\begin{equation}
[\bm a\cdot\bm\sigma,\bm b\cdot\bm\sigma]
=2i(\bm a\times\bm b)\cdot\bm\sigma.
\end{equation}

Applying the same conversion to Eq.~\eqref{eq:cqdvn-cqd-vector} gives the implicit nonlinear von Neumann equation
\begin{equation}
{
i\hbar\dot\rho
-
\hbar s k_i[\dot\rho,\rho]
=
[H,\rho].
}
\label{eq:cqdvn-implicit}
\end{equation}
Without the explicit signum factor, this reduces to Eq.~\eqref{eq:original-nonlinear-vn}, the nonlinear variant previously written for the induction equation.

\subsection{Explicit double-commutator form}

For the pure state represented by Eq.~\eqref{eq:cqdvn-rho},
\begin{equation}
\rho^2=\rho.
\end{equation}
Differentiating gives
\begin{equation}
\rho\dot\rho+\dot\rho\rho=\dot\rho,
\end{equation}
and consequently
\begin{equation}
\rho\dot\rho\rho=0.
\end{equation}
Therefore,
\begin{align}
[\rho,[\rho,\dot\rho]]
&=\rho^2\dot\rho-2\rho\dot\rho\rho+\dot\rho\rho^2\\
&=\rho\dot\rho+\dot\rho\rho\\
&=\dot\rho.
\label{eq:cqdvn-projectoridentity}
\end{align}

Rewrite Eq.~\eqref{eq:cqdvn-implicit} as
\begin{equation}
i\hbar\dot\rho+\hbar s k_i[\rho,\dot\rho]=[H,\rho].
\label{eq:cqdvn-first}
\end{equation}
Taking a commutator with \(\rho\) from the left and using Eq.~\eqref{eq:cqdvn-projectoridentity} gives
\begin{equation}
i\hbar[\rho,\dot\rho]+\hbar s k_i\dot\rho
=[\rho,[H,\rho]].
\label{eq:cqdvn-second}
\end{equation}
Eliminating \([\rho,\dot\rho]\) between Eqs.~\eqref{eq:cqdvn-first} and \eqref{eq:cqdvn-second} yields
\begin{equation}
{
\dot\rho
=
-\frac{i}{\hbar(1+s^2k_i^2)}[H,\rho]
+
\frac{s k_i}{\hbar(1+s^2k_i^2)}[\rho,[H,\rho]].
}
\label{eq:cqdvn-general}
\end{equation}
For either physical collapse branch, \(s=\pm1\), hence \(s^2=1\) and
\begin{equation}
{
\dot\rho
=
-\frac{i}{\hbar(1+k_i^2)}[H,\rho]
+
\frac{s k_i}{\hbar(1+k_i^2)}[\rho,[H,\rho]].
}
\label{eq:cqdvn-branch}
\end{equation}
Equivalently,
\begin{equation}
{
i\hbar\dot\rho
=
\frac{1}{1+k_i^2}
\left\{
[H,\rho]+i s k_i[\rho,[H,\rho]]
\right\}.
}
\label{eq:cqdvn-branchih}
\end{equation}
Setting \(k_i=0\) in Eq.~\eqref{eq:cqdvn-branchih} eliminates the induction contribution and recovers the ordinary von Neumann equation,
\[
i\hbar\dot\rho=[H,\rho].
\]
At the branching boundary \(s=0\), the general expression in Eq.~\eqref{eq:cqdvn-general} likewise reduces to the ordinary von Neumann equation, regardless of the value of \(k_i\). The latter limit must be taken in the general expression, since Eq.~\eqref{eq:cqdvn-branchih} has already used \(s^2=1\).

\subsection{Origin of the nonlinearity}

The nested commutator is
\begin{align}
[\rho,[H,\rho]]
&=\rho(H\rho-\rho H)-(H\rho-\rho H)\rho\\
&=2\rho H\rho-\rho^2H-H\rho^2.
\end{align}
For a pure state, \(\rho^2=\rho\), so
\begin{equation}
{
[\rho,[H,\rho]]
=2\rho H\rho-\rho H-H\rho.
}
\label{eq:cqdvn-expanded}
\end{equation}
The expression retains nonlinear dependence on \(\rho\) through the term \(2\rho H\rho\), even when \(H\) is independent of the state.

For \(\rho=|\psi\rangle\langle\psi|\),
\begin{equation}
\rho H\rho
=\langle\psi|H|\psi\rangle\rho
=\langle H\rangle_\psi\rho,
\end{equation}
and hence
\begin{equation}
[\rho,[H,\rho]]
=2\langle H\rangle_\psi\rho-H\rho-\rho H.
\end{equation}
The state dependence is explicit through \(\langle H\rangle_\psi\).

\newpage
For comparison,
\begin{equation}
[H,[H,\rho]]
=H^2\rho-2H\rho H+\rho H^2
\label{eq:cqdvn-linear-double}
\end{equation}
contains \(H\) twice but remains linear in \(\rho\) if \(H\) is prescribed. Thus the number of appearances of \(H\) does not determine linearity. The CQD nonlinearity arises because \(\rho\) appears more than once.

Equation~\eqref{eq:cqdvn-linear-double} also connects with the Lindblad master equation~\cite{Lindblad1976,Gorini1976}. For a prescribed Hermitian jump operator \(L=\sqrt{\kappa}H\), where \(H^\dagger=H\) and \(\kappa\geq0\), the Lindblad dissipator is
\[
\begin{aligned}
\mathcal D_L(\rho)
&=L\rho L^\dagger-\frac12\{L^\dagger L,\rho\}\\
&=\kappa H\rho H-\frac{\kappa}{2}(H^2\rho+\rho H^2)\\
&=-\frac{\kappa}{2}[H,[H,\rho]].
\end{aligned}
\]
Here, \(\{A,B\}=AB+BA\), and \(\kappa\) has units of inverse energy squared per unit time and is distinct from the dimensionless CQD induction factor \(k_i\). For time-independent \(H\), this contribution describes dephasing in the energy eigenbasis and remains linear in \(\rho\) because \(H\) is independent of the state. In contrast, the CQD induction term \([\rho,[H,\rho]]\) is nonlinear in \(\rho\). Thus, although both terms involve double commutators, the CQD term is not a standard Lindblad dissipator with state-independent jump operators.

\subsection{Closing the branching function in terms of the density operator}

The electron polar angle \(\theta_e\) need not be propagated as an independent variable because it is encoded in the density operator.  In the \(z\)-basis, Eq.~\eqref{eq:cqdvn-rho} can be written as
\begin{equation}
\rho=
\begin{pmatrix}
\cos^2(\theta_e/2) & \tfrac{1}{2}\sin\theta_e\,e^{-i\phi_e}\\
\tfrac{1}{2}\sin\theta_e\,e^{i\phi_e} & \sin^2(\theta_e/2)
\end{pmatrix}.
\end{equation}
Hence
\begin{equation}
\rho_{11}-\rho_{22}=\cos\theta_e
\end{equation}
and, equivalently,
\begin{equation}
\Tr(\rho\sigma_z)=\cos\theta_e.
\end{equation}
For \(0\leq\theta_e\leq\pi\), the electron polar angle can therefore be determined exactly from \(\rho\) as
\begin{equation}
{
\theta_e[\rho]
=
\arccos\!\left[\Tr(\rho\sigma_z)\right]
=
\arccos(\rho_{11}-\rho_{22}).
}
\label{eq:cqdvn-thetaerho}
\end{equation}
For the pure-state density operator considered here, this reconstruction is exact within the model rather than an additional dynamical equation.

For a given CQD realization, we may further assume that the nuclear polar angle is a prescribed constant,
\begin{equation}
{\theta_n=\theta_{n0}.}
\label{eq:cqdvn-thetanconstant}
\end{equation}
This is an additional modeling assumption, rather than a consequence of the Bloch-to-von-Neumann conversion itself.  Under this assumption, the CQD branching sign becomes a functional of \(\rho\):
\begin{equation}
{
s[\rho]
=
\sgn\!\left\{
\theta_{n0}
-
\arccos\!\left[\Tr(\rho\sigma_z)\right]
\right\}.
}
\label{eq:cqdvn-srho}
\end{equation}
Thus, once \(\theta_{n0}\) is specified for a realization, \(\theta_e(t)\) can be reconstructed from \(\rho(t)\), and the branching sign can be updated directly from the instantaneous density operator.

Substituting Eq.~\eqref{eq:cqdvn-srho} into Eq.~\eqref{eq:cqdvn-general} closes the CQD equation in terms of \(\rho\), \(H\), and the prescribed parameter \(\theta_{n0}\):
\begin{equation}
{
\dot\rho
=
-\frac{i}{\hbar\left[1+k_i^2s^2[\rho]\right]}[H,\rho]
+
\frac{k_i s[\rho]}{\hbar\left[1+k_i^2s^2[\rho]\right]}
[\rho,[H,\rho]].
}
\label{eq:cqdvn-closedrho}
\end{equation}
For either collapse branch, \(s[\rho]=\pm1\), this becomes
\begin{equation}
{
\dot\rho
=
-\frac{i}{\hbar(1+k_i^2)}[H,\rho]
+
\frac{k_i}{\hbar(1+k_i^2)}
\sgn\!\left\{
\theta_{n0}-\arccos\!\left[\Tr(\rho\sigma_z)\right]
\right\}
[\rho,[H,\rho]].
}
\label{eq:cqdvn-closedbranch}
\end{equation}
The state dependence is therefore present in two places: the double commutator is nonlinear in \(\rho\), and its branch-selecting coefficient is itself a discontinuous functional of \(\rho\).  At \(\theta_e=\theta_{n0}\), the adopted signum definition gives \(s=0\), for which the general form in Eq.~\eqref{eq:cqdvn-closedrho} applies.

\subsection{State-dependent effective Hamiltonian}

Equation~\eqref{eq:cqdvn-general} can be written in von Neumann form by defining
\begin{equation}
{
H_{\mathrm{eff}}[\rho,s]
=
\frac{H-i s k_i[H,\rho]}{1+s^2k_i^2}.
}
\label{eq:cqdvn-Heffgeneral}
\end{equation}
Then
\begin{equation}
{
i\hbar\dot\rho
=[H_{\mathrm{eff}}[\rho,s],\rho].
}
\label{eq:cqdvn-heffvn}
\end{equation}
For \(s=\pm1\),
\begin{equation}
{
H_{\mathrm{eff}}[\rho,s]
=
\frac{H-i s k_i[H,\rho]}{1+k_i^2}.
}
\end{equation}
If \(H\) and \(\rho\) are Hermitian and \(s,k_i\) are real, then \([H,\rho]\) is anti-Hermitian, so \(-i s k_i[H,\rho]\) is Hermitian. Therefore
\begin{equation}
H_{\mathrm{eff}}^{\dagger}=H_{\mathrm{eff}}.
\end{equation}
The effective Hamiltonian is nevertheless state dependent because it contains \(\rho\).

\subsection{Interpretation}

The CQD equation separates naturally into two contributions. The first,
\begin{equation}
-\frac{i}{\hbar(1+k_i^2)}[H,\rho],
\end{equation}
is the precessional Hamiltonian term, modified by the common factor \((1+k_i^2)^{-1}\). The second,
\begin{equation}
{
\frac{s k_i}{\hbar(1+k_i^2)}[\rho,[H,\rho]],
}
\end{equation}
is nonlinear in \(\rho\) and changes sign according to the CQD branching variable
\begin{equation}
s=\sgn(\theta_n-\theta_e).
\end{equation}
Thus the co-quantum variable selects the direction of the nonlinear induction dynamics through the signum function.

Hilbert-space representation itself remains usable: \(\rho\) and state vectors can still be represented in a Hilbert space. What is lost, in general, is a single state-independent linear propagator governing all initial states. Because \(H_{\mathrm{eff}}\) depends on \(\rho\), a linear combination of two solutions need not itself be a solution.

\subsection{Relation to the preceding derivations}

Section~\ref{sec:bloch-vn} derives the ordinary von Neumann equation from the Bloch equation. Section~\ref{sec:discussion} introduces the induction extension in Eq.~\eqref{eq:original-induction} and its nonlinear von Neumann variant in Eq.~\eqref{eq:original-nonlinear-vn}. Appendix~\ref{app:CQD} separately supplies the signum branching condition in Eq.~\eqref{eq:original-cqd-branching} and the corresponding collapse trend in Eq.~\eqref{eq:original-cqd-collapse}. The equations in the present appendix combine these two ingredients explicitly. In particular, inserting \(s=\sgn(\theta_n-\theta_e)\) into the induction term is an extension of the displayed nonlinear von Neumann equation, motivated by the CQD branching equations in the appendix; this combined operator form was not displayed in the preceding derivations.

\end{document}